\documentclass[a4paper,12pt]{article}
\usepackage[T1]{fontenc}
\usepackage{amsmath,amssymb}
\usepackage{newtxtext,newtxmath}
\usepackage{graphicx}
\usepackage[labelfont=bf,labelsep=period]{caption}
\usepackage[margin=1in]{geometry}
\usepackage[super]{cite}
\usepackage[hidelinks]{hyperref}
\graphicspath{{figs/}}
\newcommand{\VWE}{V_{\mathrm{WE}}}
\makeatletter
\renewcommand{\@biblabel}[1]{#1.}
\makeatother

\begin{document}

\begin{center}
{\Large Lock-in phase evidence for the electro-optic effect of interfacial water on ITO electrodes}\\[1.5em]Soichiro ASHIKAGA$^{\mathrm{a},*}$, Kazuaki NAKATA$^{\mathrm{a}}$, Takayoshi KOBAYASHI$^{\mathrm{b}}$, and Eiji TOKUNAGA$^{\mathrm{a},*}$\\[1.0em]
\end{center}
\noindent $^{\mathrm{a}}$Department of Physics, Tokyo University of Science, 1-3 Kagurazaka, Shinjuku-ku, Tokyo 162-8601, Japan\\
$^{\mathrm{b}}$Department of Electrophysics, National Yang Ming Chiao Tung University, Science Building III, 1001 University Road, Hsinchu 300, Taiwan\\[0.6em]$^{*}$Corresponding Authors: ashikaga.soichiro@gmail.com (S. A.), eiji@rs.tus.ac.jp (E. T.)\\[0.6em]ORCID:\\Soichiro ASHIKAGA: 0009-0001-3651-3970\\Kazuaki NAKATA: 0000-0001-7076-9956\\Takayoshi KOBAYASHI: 0000-0002-3720-239X\\Eiji TOKUNAGA: 0000-0001-8614-8898

\section*{Abstract}
Attribution to the Pockels effect of interfacial water, reported at interfaces between transparent electrodes and aqueous solutions, has relied on using an optical model to estimate the contribution of changes in the electrode's own carrier density to the electromodulation spectrum and subtracting this contribution.
In this study, we test this attribution using the phase, without subtracting the electrode contribution with an optical model.
Because the carrier response of the electrode follows charge modulation at several hundred hertz without delay, it has a common phase at all wavelengths apart from sign reversals of the response.
However, the measured transmission modulation spectra at the interface between indium tin oxide (ITO) and 0.1~mol~L$^{-1}$ aqueous NaCl show a phase peak near 420~nm, where the ITO response becomes small. This peak cannot be explained solely by a step-like change due to sign reversal of the response.
This peak appears even when the working-electrode potential, the modulation frequency, or the electrode is changed.
Furthermore, this phase peak can be explained by a thin-film calculation assuming that a uniform refractive-index change on the electrolyte side contributes with a phase different from that of the electrode response.
This result, an observation independent of the estimate of the electrode contribution, supports the attribution to the electro-optic effect of interfacial water.

\medskip
\noindent\textbf{Keywords:} Electro-optic effect, Interfacial water, Indium tin oxide electrode, Electromodulation spectroscopy

\section{Introduction}

In the electric double layer at an electrode--electrolyte interface, water molecules are subjected to electric fields reaching $10^{8}$--$10^{9}$~V~m$^{-1}$.
The response of this water to the electric field affects electrode reactions and adsorption, but methods for probing it optically in situ under an applied potential are limited.\cite{ye2009,matsui2022}
One such method is electromodulation transmission spectroscopy with transparent electrodes. At interfaces between transparent oxide electrodes such as indium tin oxide (ITO) and aqueous solutions, the Pockels effect of interfacial water, in which the refractive index of the water in the electric double layer changes in proportion to the electric field, has been reported.\cite{tokunaga2007,kanemaru2017,ashikaga2026}

The optical constants of transparent electrodes themselves also change with potential.
In degenerate ITO, changes in carrier density alter the position of the absorption edge and the free-carrier response.\cite{hamberg1984,fujiwara2005}
The observed transmission change contains overlapping contributions from the electrode and solution sides, so previous attributions have relied on estimating the electrode contribution with a model that assumes a refractive-index change in ITO, and on agreement of the remainder with a uniform refractive-index change in water.\cite{kanemaru2017}
On the other hand, the electromodulated optical response has been treated as a complex quantity in electrolyte electroreflectance\cite{gandia1988} and optical transmission.\cite{rojas2019} The phase delay at a single wavelength has also been determined from the in-phase and quadrature components,\cite{morozumi2024} and multiple simultaneous processes have been distinguished using the real and imaginary parts.\cite{nagatani2007}
If this complex treatment is used to separate the electrode and electrolyte-side contributions, it becomes possible to determine, without relying on an optical model of the electrode, whether the observed modulation can be described by the electrode response alone.

In this study, we made this determination using the criterion that the electrode carrier response has a common phase at all wavelengths apart from its sign, and examined the phase and imaginary part of complex modulation spectra at the ITO\textbar 0.1~mol~L$^{-1}$ NaCl(aq) interface.

\section{Experimental}

\subsection{Electrodes and cell}
The working electrode (WE) was an ITO film ($300 \pm 20$~nm thick) on a quartz substrate.
The cell had a three-electrode configuration similar to that reported previously.\cite{ashikaga2026} The WE and counter electrode were immersed in a 0.1~mol~L$^{-1}$ aqueous solution prepared by dissolving NaCl in purified water, and an Ag$|$AgCl reference electrode was placed in the bulk solution (Fig.~1).
Three ITO films of the same type were used as the WE and are called electrodes 1, 2, and 3.
For electrode 1, the counter electrode was a Pt rod; for electrode 2, it was an ITO electrode placed in parallel, as in the previous report;\cite{ashikaga2026} and for electrode 3, it was a Pt rod placed near the WE.
The WE was grounded, and the output of a function generator was applied to the counter electrode.
The potential of the reference electrode relative to ground was measured with an oscilloscope through an impedance buffer, and the negative of its time average was taken as the WE potential $\VWE$ vs. Ag$|$AgCl.

\begin{figure}[tbp]
\centering
\makebox[86mm][l]{\textbf{(a)}}\\
\includegraphics[width=86mm]{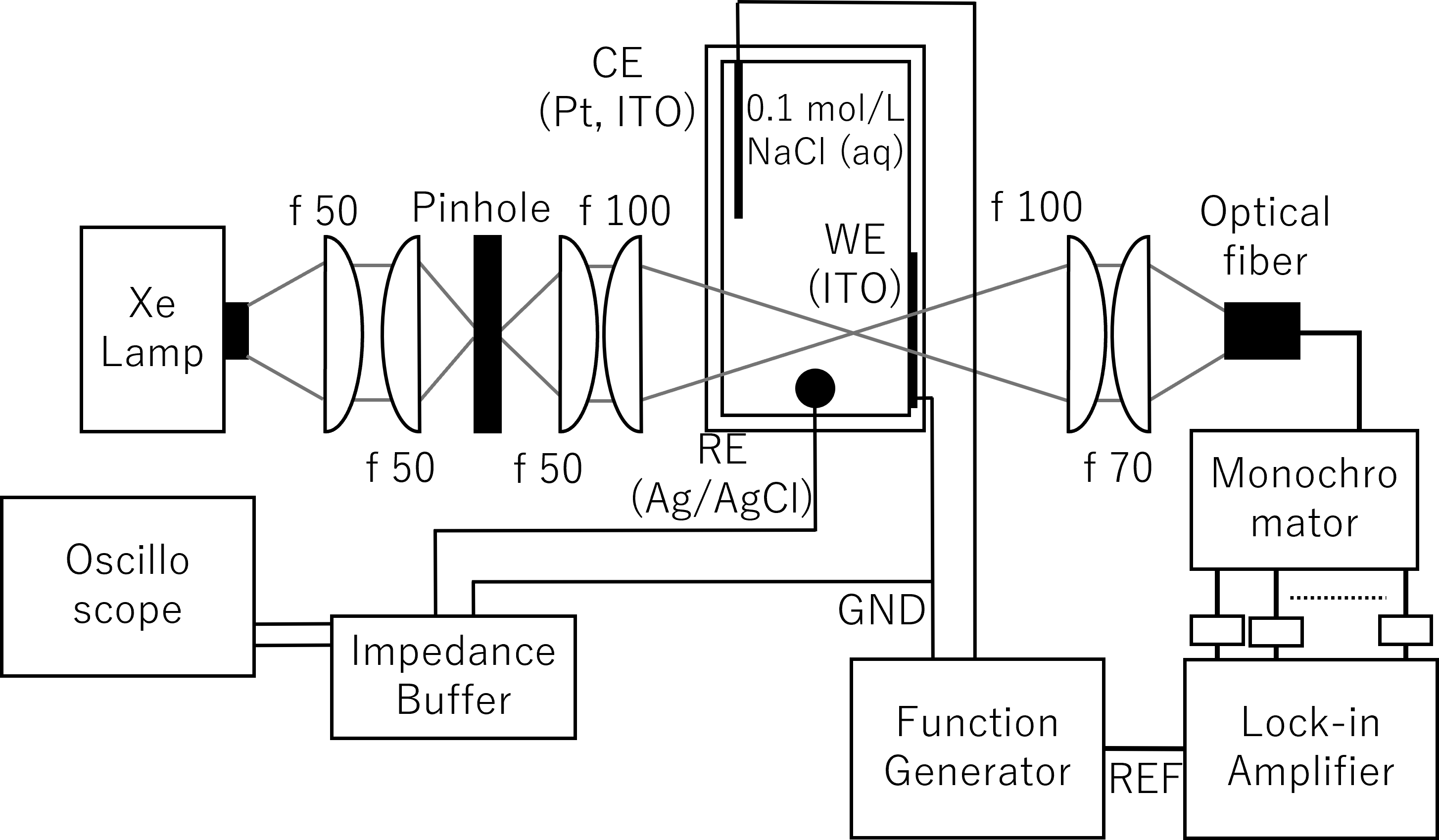}\\[1.5em]
\includegraphics[width=86mm]{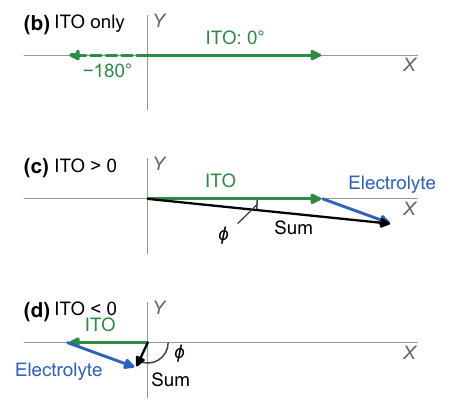}
\caption{(a) Schematic of the experimental setup.
WE: ITO; RE: Ag$|$AgCl reference electrode; CE: counter electrode (Pt rod for electrodes 1 and 3, ITO for electrode 2).
(b)--(d) Schematics of the rotated $\Delta T/T$ in the complex plane.
(b) ITO response only.
(c) Positive and (d) negative ITO responses, each with an added phase-shifted component from the electrolyte side.
In (d), the ITO response cancels the in-phase part of the electrolyte-side component, so the real part of the sum approaches zero and the phase lies between $-90^{\circ}$ and $-180^{\circ}$.}
\label{fig:1}
\end{figure}

\subsection{Electromodulation transmission spectroscopy}
The function generator output was a sine wave superimposed on a DC offset. The sine-wave frequency was 21~Hz for electrodes 1 and 2 and 26--131~Hz for electrode 3, and the AC voltage $V_{\mathrm{ac}}$ was 1.6~V$_{\mathrm{pp}}$ for electrode 1, 0.75~V$_{\mathrm{pp}}$ and 1.0~V$_{\mathrm{pp}}$ for electrode 2, and 1.8--2.0~V$_{\mathrm{pp}}$ for electrode 3.
Light from a Xe lamp was incident normally on the WE from the water side (water $\to$ ITO $\to$ quartz), and the transmitted light was dispersed with a grating spectrometer. The in-phase component $X$ and quadrature component $Y$ were detected with a 128-channel lock-in amplifier.
The complex signal $X + \mathrm{i}Y$ was normalized by the unmodulated transmitted light to give $\Delta T/T$.
The wavelength axis was calibrated using Hg lamp emission lines, and the 320--610~nm range was used for analysis. Channels with weak unmodulated transmitted light, known defective channels, and isolated outliers were excluded.

\subsection{Analysis}
Each spectrum was rotated by a single wavelength-independent phase.
The rotation phase was chosen so that, for the complex data at 320--380~nm, which are strongly affected by changes in the ITO absorption edge, the principal axis through the origin coincided with the real axis and the sum of the real parts was positive.
For spectra whose real part becomes negative between zero crossings, the phase after rotation is shown with 0° and $-180^{\circ}$ as references; for the other spectra, it is wrapped into $\pm 90^{\circ}$.

We used the transfer-matrix method for the comparison calculations.\cite{katsidis2002}
The layer stack was solution\textbar electric double layer\textbar ITO (290~nm)\textbar quartz, and the dielectric function of ITO was determined from the static transmission and reflection spectra of a separate ITO film of the same type.
We assumed that the component remaining in the imaginary part after rotation arose from a real refractive-index change $\Delta n$ that was uniform within the layer corresponding to the electric double layer (thickness 0.95~nm, the Debye length of 0.1~mol~L$^{-1}$ NaCl) and independent of wavelength. We compared its $\Delta T/T$ shape with the data using a least-squares fit over 320--610~nm with a single amplitude coefficient.
Because this thickness is much smaller than the wavelength, the calculated response is proportional to the product of $\Delta n$ and the thickness, and its shape is nearly independent of the chosen thickness.

\section{Results and discussion}

\subsection{Phase of the electrode response}
Hereafter, phase denotes the argument of the complex modulation signal detected at angular frequency $\Omega$.
The optical response accompanying the carrier-density modulation of ITO is a superposition of the shift of the absorption edge, band-gap narrowing, and the free-carrier response.\cite{hamberg1984,fujiwara2005}
Although these responses differ in speed, all occur on timescales much shorter than the modulation period at several hundred hertz and follow changes in electrode charge without delay.
The fundamental component of the transmission change is therefore in phase or in antiphase with the electrode charge at every wavelength and, for a common $\phi$, is a real number multiplied by $\mathrm{e}^{\mathrm{i}\phi}$.
Thus, when the data are rotated by this phase, the phase is expected to show a step between 0° and $-180^{\circ}$ at wavelengths where thin-film interference changes the sign of the response (Fig.~1b).
Even if the rotation phase in Section 2.3 deviates from $\phi$, only the two levels of the step shift; the phase does not take values between them.
If a component with a phase different from that of the electrode response is added, however, an imaginary part remains after rotation, and the phase takes values between the levels of the step (Figs.~1c and 1d).

\subsection{Observed phase peak}
The phase of the spectrum for electrode 1 at $\VWE = 0.433$~V is shown in black in Fig.~2c.
The phase has a peak near 420~nm and takes values from $-110^{\circ}$ to $-160^{\circ}$ over 415--426~nm.
This wavelength dependence differs from the step expected in Section 3.1.
The real part (Fig.~2a) crosses zero at 415~nm and 426~nm and reaches a minimum of $-0.19 \times 10^{-4}$ between them.
It also crosses zero at 532~nm.

\begin{figure}[tbp]
\centering
\includegraphics[width=86mm]{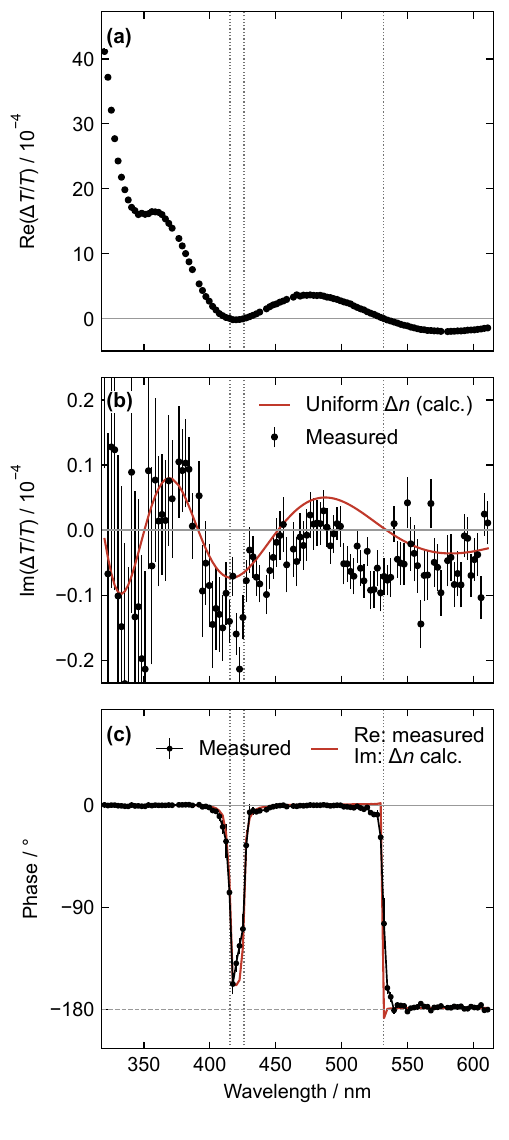}
\caption{(a) Real part, (b) imaginary part, and (c) phase of $\Delta T/T$ for electrode 1 (21~Hz, $V_{\mathrm{ac}} = 1.6$~V$_{\mathrm{pp}}$, function generator DC offset 0~V, $\VWE = 0.433$~V).
The solid line in (b) shows a thin-film calculation assuming a uniform real refractive-index change $\Delta n$ on the electrolyte side (electric double layer), with one fitted coefficient.
The red curve in (c) is the phase obtained with the measured real part and the calculated values in (b) as the imaginary part.
The error bars in (a) and (b) show the noise at each point estimated from second-order differences of adjacent channels, and those in (c) are obtained by propagating these estimates.
Vertical dotted lines mark the wavelengths at which the real part crosses zero.
A negative real response is shown at $-180^{\circ}$.}
\label{fig:2}
\end{figure}

Other spectra also show a phase peak at the same wavelength.
In spectra whose real part becomes strongly negative between zero crossings (Fig.~3a), the peak reaches nearly $-180^{\circ}$.
In spectra whose real part approaches zero near 420~nm without changing sign (Fig.~3b), there is a peak of $-20^{\circ}$ to $-50^{\circ}$ where the phase would remain at 0° if the electrode response alone were present.
For electrode 3, a peak appears at the same wavelength even when the modulation frequency is varied over 26--131~Hz (Fig.~3c).
Because the same peak also appears for electrodes 2 and 3, it is not limited to a particular electrode.
The depth and width of the peak vary with the DC potential and the frequency.
Quantifying this dependence requires an evaluation of how the charge is distributed between the electrode and the electric double layer, so it is left for future work.

\begin{figure}[tbp]
\centering
\includegraphics[width=86mm]{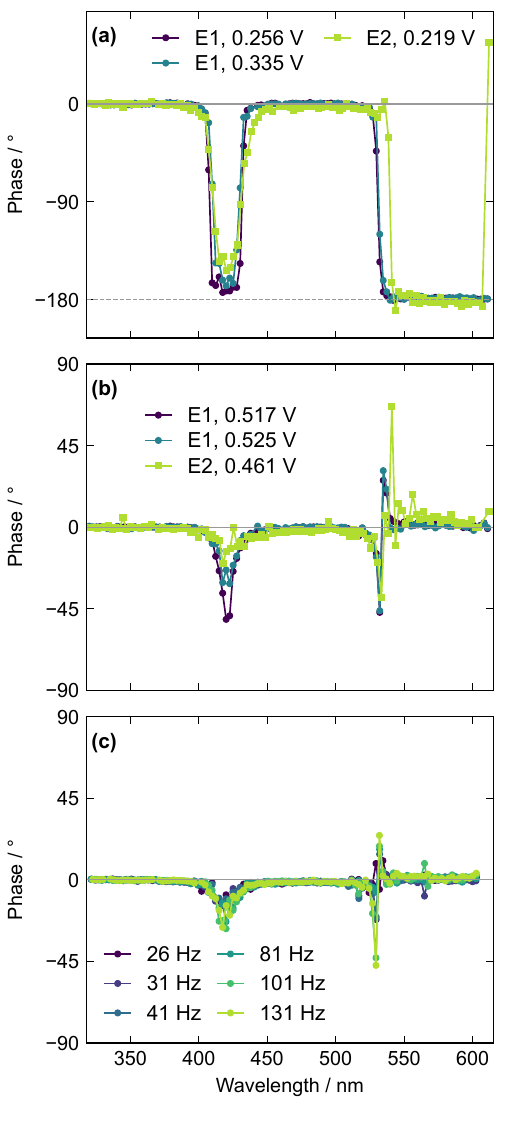}
\caption{Phase of $\Delta T/T$ for other spectra processed in the same way as in Figure~2.
(a) Spectra whose real part becomes negative between zero crossings (21~Hz, $\VWE = 0.22$--$0.34$~V), (b) spectra whose real part does not become negative (21~Hz, $\VWE = 0.46$--$0.53$~V), and (c) spectra for electrode 3 at different modulation frequencies ($\VWE = 0.44$--$0.48$~V).
E1 and E2 in the legends denote electrodes 1 and 2, respectively, and the numbers give the values of $\VWE$.
The phase in (b) and (c) is wrapped into $\pm 90^{\circ}$.
In this representation, the $180^{\circ}$ step due to sign reversal of a real response does not appear.
Changes that pass through intermediate phases remain, however, and are visible as the peak near 420~nm and the excursion near 530~nm.
The ITO films used as electrodes 2 and 3 are separate from the film used as electrode 1.}
\label{fig:3}
\end{figure}

\subsection[Uniform refractive-index change on the electrolyte side and the electro-optic effect of interfacial water]{\raggedright Uniform refractive-index change on the electrolyte side and the electro-optic effect of interfacial water}
The solid line in Fig.~2b shows the shape of $\Delta T/T$ due to a uniform $\Delta n$ on the electrolyte side.
The positions of the extrema in the imaginary part are consistent with this shape: a maximum near 370~nm, a minimum near 415~nm, and a maximum near 485~nm.
The phase obtained from the measured real part with these calculated values substituted for the imaginary part (red in Fig.~2c) reproduces the depth and width of the peak near 420~nm.
The observed phase features can be explained if a uniform refractive-index change on the electrolyte side contributes with a phase different from that of the electrode response.

Previous reports\cite{tokunaga2007,kanemaru2017,ashikaga2026} have treated the remainder after subtracting the electrode contribution with an optical model as a uniform $\Delta n$ on the electrolyte side and have attributed it to the Pockels effect of interfacial water.
In this study, using only the fact that the electrode response has the same phase at all wavelengths apart from its sign, we showed that the phase has features that cannot be explained by the electrode response alone.
Its shape is the same as that of the uniform $\Delta n$ on the electrolyte side used for the attribution in previous reports. This supports the previous attribution through an observation independent of the estimate of the electrode contribution.
Because it requires no optical model of the electrode, this phase-based test can also be applied to electromodulation spectroscopy of other transparent electrodes with a carrier response.

\section{Conclusions}
We examined the phase of modulation spectra without an optical model, using the criterion that the electrode carrier response has a common phase at all wavelengths apart from its sign.
At the ITO\textbar 0.1~mol~L$^{-1}$ NaCl(aq) interface, a phase peak that cannot be explained by a step-like change alone appeared near 420~nm, where the ITO response becomes small. Over 415--426~nm, the phase took values from $-110^{\circ}$ to $-160^{\circ}$.
This peak appears at working-electrode potentials of 0.22--0.53~V, at modulation frequencies of 21--131~Hz, and for all three electrodes.
These phase features can be explained if a uniform refractive-index change on the electrolyte side contributes with a phase different from that of the electrode response.
This phase-based test can also be applied to other transparent electrodes whose carrier response has a common phase, to distinguish refractive-index changes on the electrolyte side from the electrode contribution in electromodulation spectroscopy.

\section*{Use of Generative AI}
During the preparation of this work, the authors used Claude Code (Anthropic) and Codex (OpenAI) to assist with coding and with drafting, translating, and editing the text.
After using these tools, the authors reviewed and edited the content as needed and take full responsibility for the content of the published article.

\section*{Acknowledgments}
This work was supported by JSPS KAKENHI Grant Number JP25K01698.

\section*{Data Availability Statement}
The data that support the findings of this study are available from the corresponding author upon reasonable request.

\end{document}